\documentclass[aps,prapplied,amsmath,amssymb,twocolumn,superscriptaddress]{revtex4-2}

\usepackage{graphicx}
\usepackage{svg}
\usepackage{adjustbox}
\usepackage{fixltx2e}
\usepackage{makecell}
\usepackage{hyperref}
\usepackage{cleveref}
\usepackage{subcaption}
\usepackage{caption} 
\graphicspath{ {./pictures/} }

\begin{document}


\title{3D-Integrated Superconducting qubits: CMOS-Compatible, Wafer-Scale Processing for Flip-Chip Architectures}


\author{T. Mayer}
\email{thomas.mayer@emft.fraunhofer.de}
\affiliation{Fraunhofer Institut für Elektronische Mikrosysteme und Festkörpertechnologien EMFT, Munich, Germany}

\author{H. Bender}
\affiliation{Fraunhofer Institut für Elektronische Mikrosysteme und Festkörpertechnologien EMFT, Munich, Germany}

\author{S.J.K. Lang}
\affiliation{Fraunhofer Institut für Elektronische Mikrosysteme und Festkörpertechnologien EMFT, Munich, Germany}

\author{Z. Luo}
\affiliation{Technical University Munich, Munich, Germany}

\author{J. Weber}
\affiliation{Fraunhofer Institut für Elektronische Mikrosysteme und Festkörpertechnologien EMFT, Munich, Germany}

\author{C. Morán Guizán}
\affiliation{Fraunhofer Institut für Elektronische Mikrosysteme und Festkörpertechnologien EMFT, Munich, Germany}

\author{C. Dhieb}
\affiliation{Fraunhofer Institut für Elektronische Mikrosysteme und Festkörpertechnologien EMFT, Munich, Germany}

\author{D. Zahn}
\affiliation{Fraunhofer Institut für Elektronische Mikrosysteme und Festkörpertechnologien EMFT, Munich, Germany}

\author{L. Schwarzenbach}
\affiliation{Fraunhofer Institut für Elektronische Mikrosysteme und Festkörpertechnologien EMFT, Munich, Germany}

\author{W. Hell}
\affiliation{Fraunhofer Institut für Elektronische Mikrosysteme und Festkörpertechnologien EMFT, Munich, Germany}

\author{M. Andronic}
\affiliation{Fraunhofer Institut für Elektronische Mikrosysteme und Festkörpertechnologien EMFT, Munich, Germany}

\author{A. Drost}
\affiliation{Fraunhofer Institut für Elektronische Mikrosysteme und Festkörpertechnologien EMFT, Munich, Germany}

\author{K. Neumeier}
\affiliation{Fraunhofer Institut für Elektronische Mikrosysteme und Festkörpertechnologien EMFT, Munich, Germany}

\author{W. Lerch}
\affiliation{Fraunhofer Institut für Elektronische Mikrosysteme und Festkörpertechnologien EMFT, Munich, Germany}

\author{L. Nebrich}
\affiliation{Fraunhofer Institut für Elektronische Mikrosysteme und Festkörpertechnologien EMFT, Munich, Germany}

\author{A. Hagelauer}
\affiliation{Fraunhofer Institut für Elektronische Mikrosysteme und Festkörpertechnologien EMFT, Munich, Germany}
\affiliation{Technical University Munich, Munich, Germany}

\author{I. Eisele}
\affiliation{Fraunhofer Institut für Elektronische Mikrosysteme und Festkörpertechnologien EMFT, Munich, Germany}
\affiliation{Center Integrated Sensor Systems (SENS), Universität der Bundeswehr München, Munich, Germany}

\author{R.N. Pereira}
\affiliation{Fraunhofer Institut für Elektronische Mikrosysteme und Festkörpertechnologien EMFT, Munich, Germany}

\author{C. Kutter}
\affiliation{Fraunhofer Institut für Elektronische Mikrosysteme und Festkörpertechnologien EMFT, Munich, Germany}
\affiliation{Center Integrated Sensor Systems (SENS), Universität der Bundeswehr München, Munich, Germany}

\email{thomas.mayer@emft.fraunhofer.de}



\date{\today}

\begin{abstract}
In this article, we present a technology development of a superconducting qubit device 3D-integrated by flip-chip-bonding and processed following CMOS fabrication standards and contamination rules on 200\,mm wafers. We present the utilized proof-of-concept chip designs for qubit- and carrier chip, as well as the respective front-end and back-end fabrication techniques. In characterization of the newly developed microbump technology based on metallized KOH-etched Si-islands, we observe a superconducting transition of the used metal stacks and radio frequency (RF) signal transfer through the bump connection with negligible attenuation. In time-domain spectroscopy of the qubits we find high yield qubit excitation with energy relaxation times of up to 15\,µs.
\end{abstract}


\maketitle

\section{Introduction}
The superconducting qubit architecture holds multiple advantages over other quantum computing platforms the make it a leading candidate for the realization of general-purpose, fault-tolerant quantum computing: fast gate speeds\cite{doi:10.1126/science.1231930}, relatively long coherence times\cite{Wendin_2017}, and especially the possibility to manufacture quantum processing units (QPUs) with precisely engineered properties by macroscopic, lithographically-defined nanofabrication. However, while dramatic progress has been demonstrated recently \cite{Biznarova2024-dj, Bal2024-sz, Google_Quantum_AI_and_Collaborators2025-hv}, still a significant scaling towards much larger qubit numbers will be necessary and hence some crucial engineering challenges are yet to be solved\cite{10.1063/1.5089550}. One of these challenges is the high-precision fabrication of Josephson Junction (JJ) that is necessary to not only reach the required qubit performance, but also the targeting of critical device parameters like the qubit frequency\cite{PhysRevResearch.5.043001}. The latter is especially relevant with increasing chip sizes of QPU devices hosting many qubits, where the processing accuracy of the commonly applied shadow-evaporation and lift-off method for JJ processing could run into limitations due to inherent on-wafer gradients of the fabrication technique. As an alternative to this processing approach, the utilization of CMOS industry-style tooling and methods has played an increasingly prominent role in the community recently, as it promises high precision device fabrication on large wafer scales ($\geq 200$\,mm) and superior repeatability and process control and monitoring\cite{Van_Damme2024-hz, Verjauw2022-bo}. A second challenge comes with the necessity of 3D-integration of superconducting QPU devices when scaling to a large number of qubits\cite{Rosenberg2017-hv, Yost2020-si, Foxen_2018}. This is, for example, crucial to enable uniform grounding of the superconducting base layer across the entire chip area and indispensable to facilitate the routing of all required signal and control lines to operate the device.\\
In this work, we present an approach that addresses both of these challenges. We fabricate a proof-of-concept QPU device hosting 24 floating, fixed-frequency qubits in a foundry-style 200\,mm clean room using commercial equipment and methods from the semiconductor industry following CMOS contamination standards. Using the same infrastructure, carrier devices are manufactured on separate 200mm wafers that host a uniquely developed microbump technology for flip-chip-bonding. In the final, chip-to-chip bonded devices, the microbump connections are used to facilitate uniform grounding and to transfer the radio frequency (RF) signal between qubit and carrier-chip. The microbumps are characterized by superconductivity and RF-transmission measurements. By performing spectral analysis and time-domain measurements on the qubits, we demonstrate successful qubit excitation with high yield and good coherence times for the first processing run.




\section{Chip design}\label{ch:chip_design}

The chip designs used to demonstrate our technologies are presented in Fig. \ref{fig:chip_design}. The final 3D-integrated superconducting qubit device consists of a qubit and carrier chip that are fabricated on separate 200\,mm wafers (see Sec. \ref{FE_fab}) and mechanically and electrically connected by flip-chip bonding. The microbump connections between the two chips are utilized to facilitate homogeneous grounding of the qubit chip's base layer and to transfer RF- control and readout signals between the two chips. A more in-depth description of the chip design and circuit modelling details can be found in Ref. \cite{10974912}.


\subsection{Qubit chip design}
\begin{figure*}[]
    \centering
    \includegraphics[width=\textwidth]{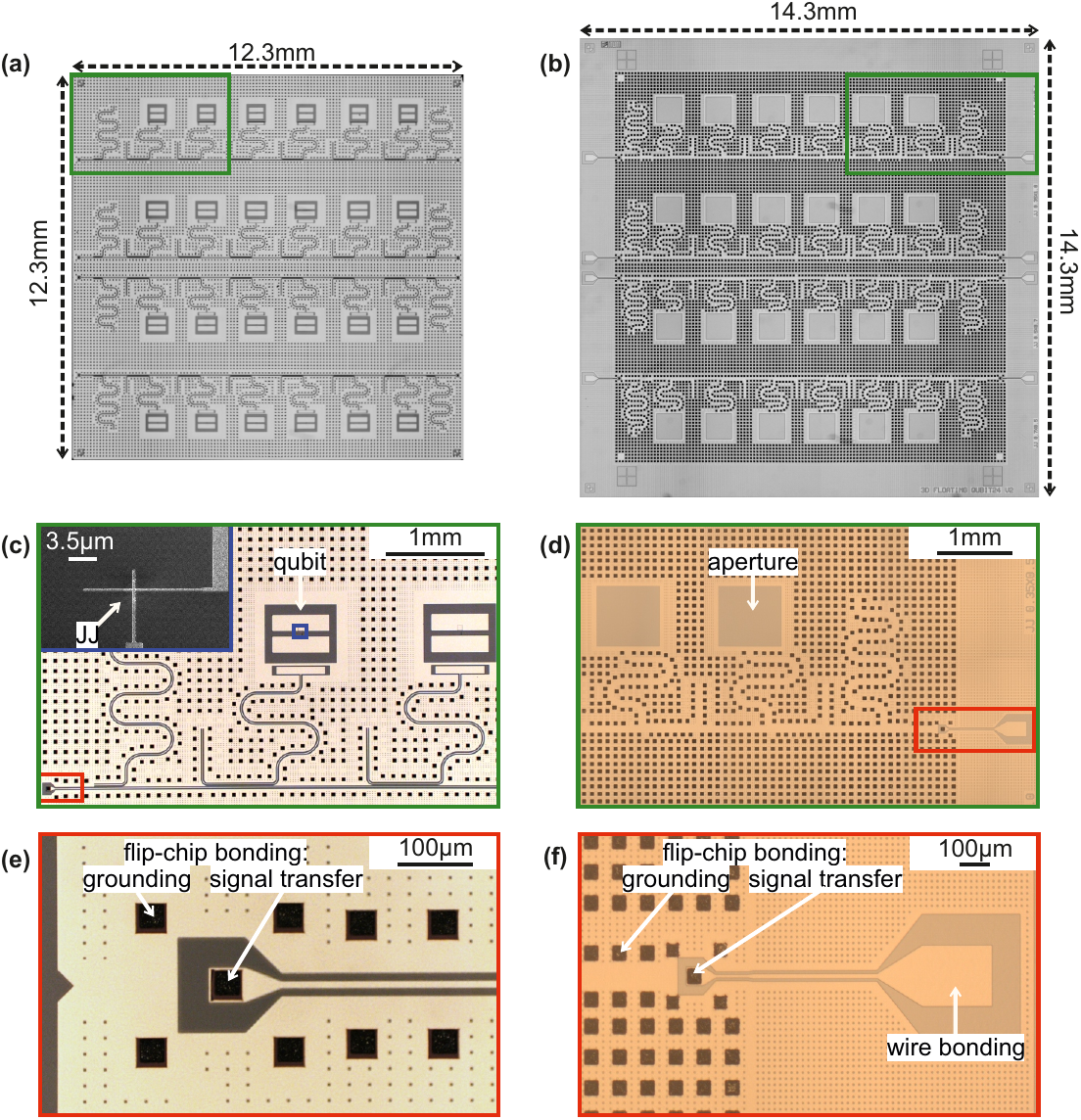}
    \caption{Micrograph of qubit chip (a) and carrier chip (b). Larger magnification images of the respective chips c) - f).} 
    \label{fig:chip_design}
\end{figure*}

The qubit chip (Fig. \ref{fig:chip_design}a)) is quadratic with an edge length of 12.3\,mm and hosts 24 fixed-frequency qubits spaced across four separate feedlines. The qubits are designed in a typical pocket Transmon architecture, where a Josephson Junction is placed between two isolated plates of a shunt capacitor to reduce the sensitivity of the system to charge noise. Figure \ref{fig:chip_design}c) shows a close-up featuring two qubits, a reference resonator, and part of the feedline surrounded by the superconducting base-layer. The base-layer, feedline, resonators and shunt capacitors consist of Aluminum, the Josephson Junction highlighted in the inset of Fig. \ref{fig:chip_design}c) of an Al/AlOx/Al stack. Spread across the base layer, holes through the Al to the Si substrate are patterned to avoid vortices\cite{Chiaro2016-ot}. In-between, pads are metallized by a TiN/In stack to facilitate the superconducting connection to the carrier chip. The pads have a size of 37$\times$37\,µm$^2$ and typical pitch of 100\,µm for optimal homogeneous grounding across the device. Importantly, one pad is placed on each end of the feedlines to allow transmission of the radio-frequency (RF) signal from the carrier to the qubit chips (as shown in Fig. \ref{fig:chip_design}e)).

\subsection{Carrier chip design}
The carrier chip (Fig. \ref{fig:chip_design}b)) is larger than the qubit chip with an edge length of 14.3\,mm as it additionally hosts the pads for wire-bonding the final device to a cryo-compatible printed circuit board. The close-up micrograph shown in Fig. \ref{fig:chip_design}d) represents the counterpart area to the qubit chip presented in Fig. \ref{fig:chip_design}c). The chip is metallized by a Niobium layer and patterned with the necessary signal transfer wiring and apertures with carefully engineered window size. In the final flip-chip bonded device, these apertures are located directly opposite the qubits, as the removal of the metal on the carrier chip in the vicinity of the qubits aims to reduce dielectric losses due to the lossy metal/air interface\cite{PhysRevApplied.21.054063}.
The microbumps for flip-chip bonding consist of Si-island metallized by a Nb/In stack placed across the chip to directly attach to the TiN/In pads on the qubit chip in the bonding process. Figure \ref{fig:chip_design}f) again shows a micrograph close-up referring to the same area on the chip shown in Fig. \ref{fig:chip_design}e). In contrast to the qubit chip, the carrier chip hosts additional large pads at its edges for wire bonding, which is used to connect the device to the outside electronics. The RF-signal applied to control and characterize the final 3D integrated qubit chip is transferred to the device via the wire bonds before entering and exiting the qubit chip through the  microbump connection on the feedline.

\section{Front-end fabrication}\label{FE_fab}
\begin{figure*}[!htb]
    \centering
    \includegraphics[width=\textwidth]{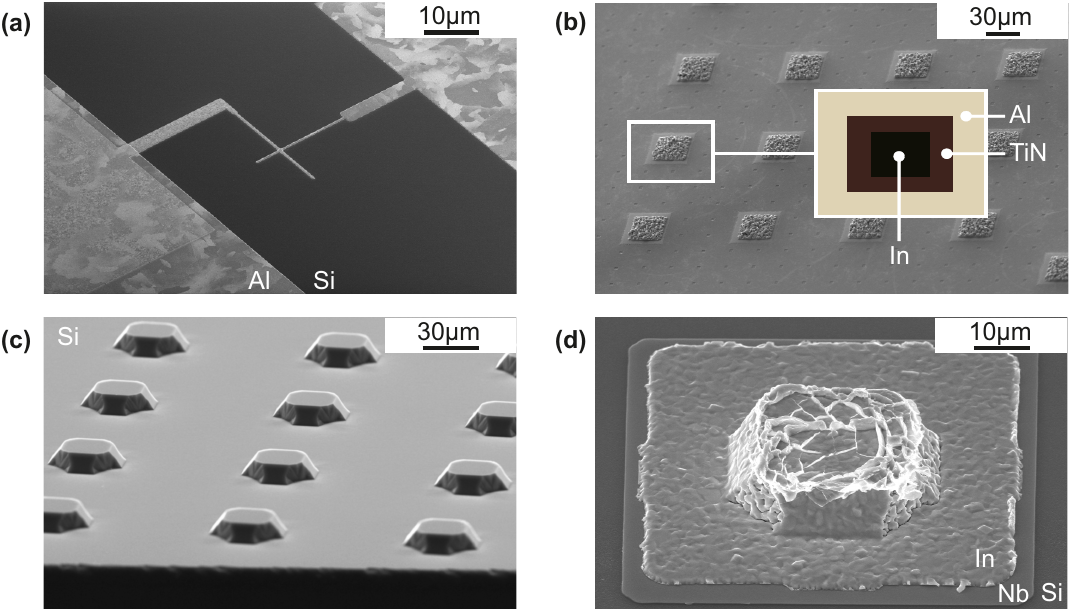}
    \caption{Scanning Electron Microscopy images of various processing stages. Finished Josephson Junction (a), metallization pads for bump connection on the qubit chip (b), KOH-etched Si-islands on the carrier chip (c), metallized Si-islands for superconducting microbump connection (d).}
    \label{fig:FE_processing}
\end{figure*}
The front-end fabrication of qubit and carrier chips is conducted on separate 200\,mm wafers within the pilot-line facilities of Fraunhofer EMFT in an ISO3 level clean room environment. The line is subdivided into two parts: a CMOS line, where strict CMOS contamination guidelines are followed and a separate part, where the handling of non-CMOS materials, like Niobium or Indium, is enabled. For the complete fabrication of both wafers a foundry-style processing is applied, exclusively utilizing methods established in the semiconductor industry. For example, patterning of the Josephson Junction is executed by subsequent subtractive etching of the two electrodes in two separate lithography layers instead of the double-angle evaporation and lift-off methods used in smaller scale fabrications and no airbridges are utilized for uniform grounding.
\subsection{Qubit wafer fabrication}
The processing of the qubit chip wafers starts in the CMOS line with  pristine p-type, (100) Silicon substrates of 3-5\,k$\Omega$cm resistivity to minimize substrate losses. To remove the native oxide from the substrate and hence yield an optimal metal-substrate interface, the wafer is HF-dipped before being mounted to an ultra-vacuum metal sputtering tool with a strict time coupling. A film of 150\,nm Aluminum is sputtered at room temperature as the base metallization layer of the device. Using standard i-line (365\,nm) stepper lithography, this base layer is patterned with the feedline and wire-bond pads for the signal transfer, vortex-pinning holes across the ground planes, the control resonators, both plates of the qubits’ shunt capacitors, and the bottom electrode (BE) of the Josephson Junction. The BE has a design width of 350\,nm, approaching the physical resolution limitation of the lithography tool. The patterning is done by plasma dry-etching using a Cl-based chemistry and in-situ resist ashing in an oxygen plasma to prevent Aluminum corrosion when exposed to ambient air. Since vacuum-breaking is necessary between the described steps, the forming of a native oxide on the Al surface is  unavoidable. To create a defined tunneling oxide of the Al/AlOx/Al Josephson Junction, the subsequent processing steps are hence executed in-situ: first, the native oxide is removed by Ar backsputtering before the tunneling oxide is formed via static oxidation of the bottom electrode by finely controlling the oxygen pressure and the oxidation time in the chamber. This tunneling oxide is then enclosed by depositing the second Aluminum layer (100\,nm) on the wafer. Now, the vacuum can be broken for the following second optical lithography and the patterning of the JJ’s top electrode by a similar dry etch process as used in the first layer. A SEM image of a finalized JJ is depicted in Fig. \ref{fig:FE_processing}a). More details of the fabrication process as well as reporting on the performance of planar qubit devices fabricated by this industry-style, subtractive patterning approach wcan be found in Ref. \cite{lang2025advancingsuperconductingqubitscmoscompatible}.
To facilitate the flip-chip bonding process, the qubit wafers have to undergo further processing to pattern the under-bump metallization after fabrication of the two Aluminum layers. For this metallization, a stack of TiN and In is used on top of the Aluminum base-layer. Especially as the microbump technology in our chip design is not only used for grounding, but also for RF-signal transfer between qubit- and carrier chip, all interfaces between the metal layers need to be optimized to prevent signal attenuation. At the same time, the qubits and resonators are already fabricated on the wafers and any subsequent processing can possibly harm the sensitive Aluminum surfaces and hence deteriorate qubit performance.
For the proof-of-concept presented in this work, we implemented a lift-off process for both the TiN and In layer. The lift-off procedure comes with the upside of the delicate Al surfaces being protected by the used resist that is opened only in those areas, where the under-bump metallization is formed. A possible downside lies in the necessity to prepare the respective surface in-situ prior to the metal deposition by Ar-backsputtering to remove e.g. surface layer oxides. When the lift-off resist is exposed to the ion bombardment of this step, it can be degraded, deformed, or hardened, impacting the lift-off quality and increasing the risk of resist residues on the Aluminum surfaces, which in turn cause dielectric defects in the cryogenic, high frequency characterization. Hence, a fine trade-off between sufficient surface preparation without lift-off resist degradation has to be carefully tuned for the backsputtering step. For the TiN layer, the resist is exposed again by i-line stepper lithography to open squares across the Al base layer of the chip (see section 1). After the Ar backsputtering step, a film of 40\,nm TiN is sputtered at a deposition temperature of 60°C. The lift-off is performed in an Acetone ultrasonic bath before the wafers are cleaned by a Isopropyl and DI water rinse. For the Indium layer, the lift-off sequence is repeated, however, since Indium is not considered CMOS-compatible, the wafers exit the CMOS line for further processing. The resist is exposed by a mask-aligner lithography, which is sufficient for the window sizes of 29$\times$29\,µm$^2$. The Indium squares are designed with a smaller edge length than for TiN to account for lateral offsets in the lithography steps. After Ar backsputtering, a 2\,µm thick Indium film is sputtered and the lift-off performed. The final Al/TiN/In pads are shown in the SEM image of Fig. \ref{fig:FE_processing}b).

\subsection{Carrier wafer fabrication}
The fabrication of the wafers for the carrier-chip is initiated with the identical high-ohmic, p-type Si substrate, and starts by the patterning of dedicated silicon islands (Fig \ref{fig:FE_processing}c)) that are later metallized to form the microbumps for flip-chip bonding. This Si-island technology aims to minimize the tilt and create a highly accurate and reproducible spacing between carrier- and qubit-chip, which are crucial to allow a precise targeting of electrical circuit design specifications of the device. To pattern the Si-island, a silicon-oxide hard-mask is created by thermal oxidation of the wafer, i-line stepper lithography and plasma dry-etching. Subsequently, the wafers leave the CMOS line and the islands are formed by wet-chemical etching in a KOH solution. For the metallization, a 300\,nm thick layer of Niobium is sputtered and patterned by a mask-aligner contact-lithography and dry etching. After a BOE clean to remove oxides from the Niobium surface, the islands are additionally covered with a 2.5\,µm thick Indium film applied by the identical lift-off sequence as described in the previous section. A silicon island covered by the Nb and In metallization layers is depicted in Fig. \ref{fig:FE_processing}d).

\section{Back-end fabrication}

\begin{figure}[]
    \centering
    \includegraphics[width=\linewidth]{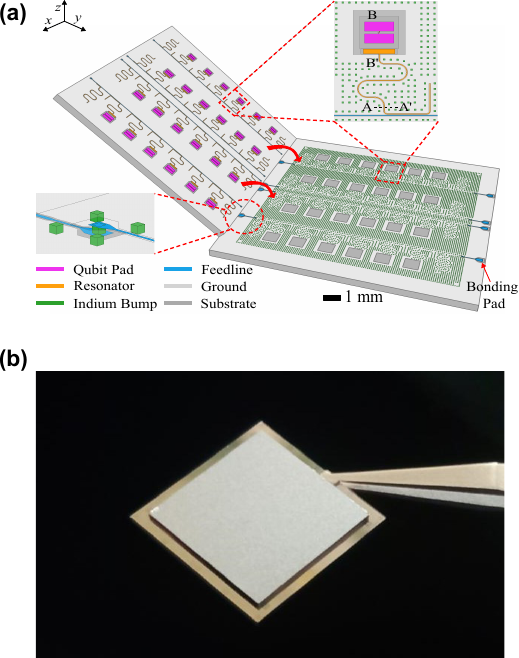}
    \caption{Schematic drawing visualizing the chip-to-chip bonding process (a) and photograph of final device (b).}
    \label{fig:fcb_schematic}
\end{figure}

After finalizing the front-end fabrication, both qubit and carrier wafers are coated with a protection resist and diced by a diamond saw into single chips. Each qubit wafer contains 151 of the 24-qubit devices presented in Section \ref{ch:chip_design}, each carrier wafer 119 chips, respectively. From those, single devices are picked and the protection resist is removed by an Acetone/Isopropyl/DI water cleaning. For the chip-to-chip bonding process a Panasonic FCB3 narrow-pitch bonder is used. Figure \ref{fig:fcb_schematic} a) shows a schematic of the alignment between the two chips in the bonding process. As shown in the respective insets, the devices are placed such that their feedlines are connected via a microbump on each side and the apertures on the carrier lie directly opposite to the qubits. A photograph of a final flip-chip-bonded device is displayed in Fig. \ref{fig:fcb_schematic} b). Lastly, the device is wire-bonded with Aluminum wires to cryo-compatible printed circuit boards (PCBs), where two wires are attached to each side of the chip's four feedlines for impedance matching and additional wires are placed all around the chip to mass for homogeneous grounding.

\section{Characterization}
\subsection{Bump characterization}

\begin{figure}[b]
    \centering
    \includegraphics[width=\linewidth]{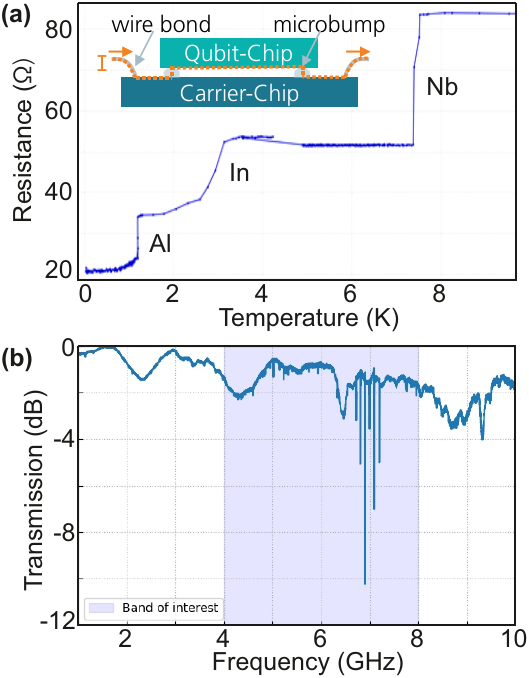}
    \caption{Measurements of superconductivity of the microbump connection using the current path as indicated in the inset (a). RF- transmission measurements of the microbump connection using the same current path (b).}
    \label{fig:4_bumpCharac}
\end{figure}

\begin{figure*}[]
    \centering
    \includegraphics[width=\textwidth]{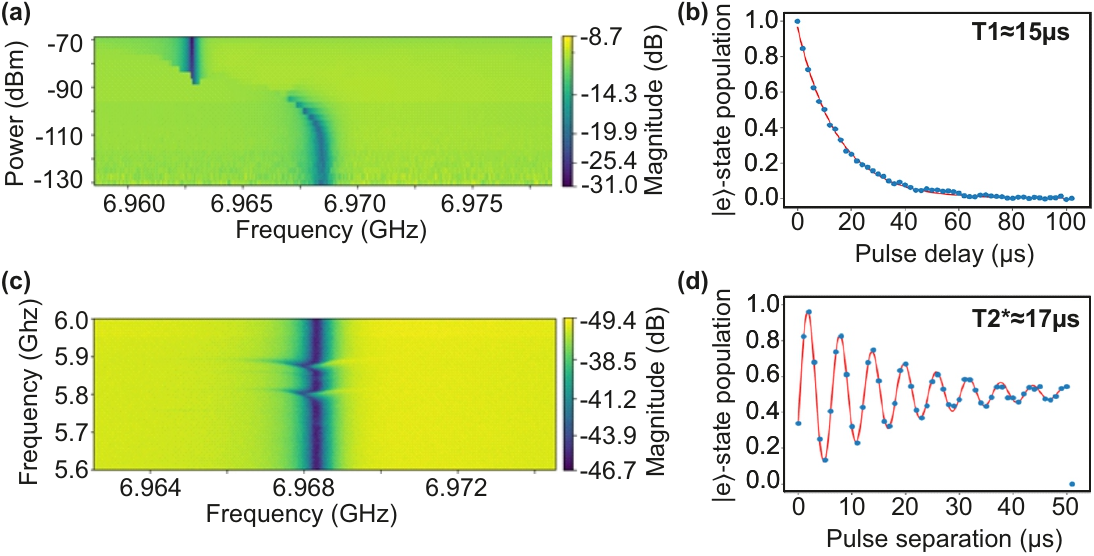}
    \caption{Exemplary measurements for one-tone (a) and two-tone (b) spectroscopy and for $T_1$ (c) and $T_2^*$ (d).}
    \label{fig:qubitperformance}
\end{figure*}

In the proof-of-concept (POC) test-chip design presented in this work, the Si-island microbump 3D integration provides two-fold functionality: on the one hand, the large number of metallized microbumps spread across the chip area aim to provide an effective and highly homogeneous grounding across the device. On the other hand, one microbump connects the qubit chips’ signal line on each side to the carrier chip for the transfer of RF-signals between the qubits and outside electronics via the Al wire-bonds and the PCB. Importantly, this RF-transfer could also be used for signal crossings on a more complex QPU design. Efficient grounding and RF-signal routing capabilities in such a 3D architecture would eliminate the need for air bridges, which are often used in planar superconducting qubit devices.
For both functionalities, the quality of the connection between the bumps on the carrier and the pads on the qubit chip is crucial. To characterize this, we measure the DC resistance of a sample as a function of temperature during a cooldown in a dilution refrigerator with a base temperature of $\approx$10\,mK (Bluefors LD 400). The feedline pads of the carrier chip are connected to the PCB via two wirebonds each, enabling a four-point measurement. The current is applied to the carrier chip, flows into the qubit chip via a microbump, passes through the $\approx$11.7\,mm aluminum feedline, and exits through another microbump towards the output port of the carrier. A Source Measurement Unit (SMU) performs a current-voltage sweep, from which the resistance is calculated. Figure \ref{fig:4_bumpCharac} a) shows the measurement results, where the superconducting transitions of Nb, In, and Al are clearly visible and close to the critical temperature values reported in the literature.


For the TiN layer used in the pad metallization stack on the qubit chip, no step corresponding to a specific $T_c$ value can be observed, which could stem from the layer not entering the superconducting state. This could also explain the remaining resistance of about 20\,$\Omega$ at $T\approx$10\,mK, which indicates an imperfect superconductivity of the current path. We consider a more likely explanation, however, that the TiN $T_c$ step is not observed due to the very small thickness and hence minimal contribution to the total current path of the layer. The remaining low temperature resistance might  stem from imperfect interfaces in the metal stack, e.g. a remaining AlOx layer at the Al/TiN interface.\newline
To allow signal exchange between qubit- and carrier chip in our POC test device and to assess this functionality for future, more complex signal routing applications, the RF-properties pose another crucial characteristic of the bump technology. To measure the RF-performance of the feedline, we apply a two-port S-parameter measurement with a Vector Network Analyzer (VNA). To calibrate the setup, we use cryogenic RF switches and mechanical standards to perform a short-open-load-thru (SOLT) calibration, effectively shifting the reference plane from the VNA output to the mK stage of the cryostat. Figure \ref{fig:4_bumpCharac}b) shows the calibrated S21 spectrum. In the relevant frequency range of 4–8\,GHz (shaded region), which hosts the six qubit resonator modes, the feedline presents low-loss signal transmission. Since the transmission characteristics are comparable to standard planar CPW signal lines, the microbump connection in the signal chain is not introducing additional insertion loss.




\subsection{Qubit characterization}

To investigate the qubit performance and reliability of the final devices, 6 signal lines across 4 different 3D assemblies were wire bonded and the devices mounted to a \textit{Bluefors LD400} dilution refrigerator. With in total 6 transmission lines connected, a maximum of 36 functioning resonator/qubit pairs can be expected. By performing initial spectral analysis the frequencies of 31/36 resonators could be successfully determined. The functionality of the 31 possible qubits was tested by performing one-tone and two-tone spectroscopy. The characteristic "punch-out" signature in the one-tone measurement (Fig. \ref{fig:qubitperformance} a)) indicates an active qubit being coupled to the respective resonator. From the two-tone assessment (Fig. \ref{fig:qubitperformance} c)) qubit frequencies and anharmonicities can be determined. Of the remaining 31 possible qubits, 29 showed the characteristic signatures in both spectroscopy methods, indicating a high fabrication yield.
The qubit performance was tested by time-domain spectroscopy to obtain values for $T_1$ and $T_2^*$ coherence times. Exemplary measurements are shown in Fig. \ref{fig:qubitperformance} b) for $T_1$ and in Fig. \ref{fig:qubitperformance} d) for $T_2^*$. With the respective values reaching 15\,µs and 17\,µs, we underline the feasibility of processing high fidelity superconducting qubits in a flip-chip architecture with our large-wafer, CMOS-compatible processing approach.

\section{Conclusion}
In this work, we presented a proof-of-concept technology development of a superconducting qubit device 3D-integrated via flip-chip bonding and discussed the microbump and qubit characteristics of a very first processing run. The applied CMOS-style front-end fabrication within the 200\,mm pilot line at Fraunhofer EMFT allows the leveraging of an industry-grade infrastructure and its high precision methods and control mechanisms, which will be crucial for future scaling towards larger and more complex devices. The newly developed microbump technology using metallized KOH-etched Si-islands are characterized via superconductivity and low-temperature RF-transmission measurements. While no full superconductivity of the bumps is observed showing the need for further process optimization, the technology still allows efficient RF-signal transfer between the two chips. Within the qubits spectroscopy analysis we observed high yields and good qubit performance promising substantial future improvement with further process development.\\
3D-integration technology will be particularly indispensable in more complex QPU devices, where the number of qubits raises the need for large-area grounding, additional routing layers, and the possibility of signal crossings. Furthermore, especially considering the large number of chips on every 200\,mm wafer, the technology also provides high potential to be used in modular superconducting QPU devices, where more than one qubit chip is flip-chip-bonded to the same carrier \cite{Gold2021-zs,10.1109/MICRO56248.2022.00078,Niu2023-yp,PhysRevApplied.21.054063}.

\section*{Acknowledgements}
The authors acknowledge helpful discussions and assistance with test chip design and setting up of cryogenic measurements from G. Huber, I. Tsitsilin, F. Haslbeck and C. Schneider from the Quantum Computing group at the Walther Meissner Institute. We would like to thank Z. Tianmu and L. Sigl from Zurich Instruments for their support in setting up the Qubit characterisation software. Moreover, we acknowledge J. Hepp and H. Adel from Fraunhofer IIS for providing us with packages and pcbs for cryo testing.\newline
We would like to thank the Fraunhofer EMFT process engineers and clean room staff for valuable discussion on process development and the professional wafer fabrication.\newline
This work was funded by the Munich Quantum Valley (MQV) – Consortium Scalable Hardware and Systems Engineering (SHARE), funded by the Bavarian State Government with funds from the Hightech Agenda Bavaria and the Munich Quantum Valley Quantum Computer Demonstrator - Superconducting Qubits (MUNIQC-SC) 13N16188, funded by the Federal Ministry of Education and Research, Germany.

\bibliography{paper.bib}
\end{document}